\documentclass[12pt,a4paper,twoside,showpacs,showkeys,nofootinbib]{revtex4}
\usepackage{amsmath,amsfonts}
\usepackage{amsbsy}
\usepackage{mathrsfs}
\usepackage{color}
\usepackage{psfrag}
\usepackage{enumerate}
\usepackage{amsmath,amssymb,calc,amsfonts}
\usepackage{latexsym}
\def\ut#1{\rlap{\lower1ex\hbox{$\sim$}}#1{}}
\DeclareFontFamily{U}{rsfs}{}         
\DeclareFontShape{U}{rsfs}{m}{n}{<5> rsfs5 <6><7> rsfs7          %
  <8><9><10><10.95><12><14.4><17.28><20.74><24.88> rsfs10}{}     %
\DeclareMathAlphabet{\mathfs}{U}{rsfs}{m}{n}                     %
\newcommand{\mfs}[1]{\mathfs {#1}}                               %
\def\beq{\begin{eqnarray}}
\def\eeq{\end{eqnarray}}
\def\nn{\nonumber\\}

\def\del{\partial}
\def\grad{\nabla}

\def\lie{\mfs L}

\begin{document}

\title{Estimation of temperature of cosmological apparent horizons : a new approach}

\author{Bhramar Chatterjee}\email{bhramar.chatterjee@gmail.com}
\author{Narayan Banerjee }\email{narayan@iiserkol.ac.in}
\affiliation{Indian Institute of Science Education and Research Kolkata, Mohanpur, Nadia 741246, INDIA.}

\begin{abstract}
 We consider radiation from cosmological apparent horizon in Friedmann-Lemaitre-Robertson-Walker(FLRW) model in a double-null coordinate setting. As the spacetime is dynamic, there is no timelike Killing vector, instead we have Kodama vector which acts as dynamical time. We construct the positive frequency modes of the Kodama vector across the horizon. The conditional probability that a signal reaches the central observer when it is crossing from the outside gives the temperature associated with the horizon.
\end{abstract}

\keywords{Apparent horizon; horizon temperature; Kodama vector.}

\pacs{04.70.Dy; 98.80.-k}

\maketitle

\section{Introduction}
Horizon thermodynamics is an extensively studied subject which acts as a window between classical and quantum theories of gravity. The laws of black hole mechanics \cite{Bardeen, Bekenstein} first gave an indication that something other than classical general relativity is at work where a horizon exists. Then Hawking discovered \cite{Hawking} that a black hole horizon indeed emits thermal radiation at a temperature $\kappa$, it's surface gravity and what was thought as a mere coincidence was now understood as the beginning of a new paradigm in physics. Soon after the discovery of black hole radiation, Gibbons and Hawking \cite{Gibbons} showed that a de Sitter cosmological horizon also emits a radiation.

In all these calculations the horizon concerned was the event horizon, a concept which is essentially global and teleological, associated with only a stationary spacetime. But black holes are subject to dynamical processes such as their formation and evolution, accretion of matter/ energy, their interaction with other objects and their own back-reaction to the emission of Hawking radiation. So the idea of event horizon is not very useful in reality\cite{Ashtekar1, Faraoni1}.

The concepts of dynamical local horizons are in fact more appropriate to deal with the dynamical aspects of black hole mechanics. Local horizons can be located and described by local geometrical properties of the embedding spacetimes without any assumptions about the global evolution. There are several examples of local horizons viz. trapping horizon\cite{Hayward1, Hayward2}, slowly evolving horizon\cite{Booth}, dynamical horizon\cite{Ashtekar1, Krishnan}, isolated horizon\cite{Ashtekar2} etc. All these horizons are endowed with nice properties i.e they obey the laws of black hole mechanics and also a temperature and an entropy can be associated with them. Although these formulations were primarily motivated by providing a realistic description of a black hole, these can be applied to other situations also, particularly in cosmology.

Radiation from cosmological horizon is also a much studied subject like black hole. Gibbons and Hawking\cite{Gibbons} considered a de Sitter cosmological model which is almost analogous to Schwarzschild black hole case since the spacetime admits a timelike Killing vector and thus admitting no genuine time evolution. For a more general Friedmann-Lemaitre-Robertson-Walker (FLRW) model, though the event horizon and particle horizon can be defined, there is no timelike Killing vector since the spacetime is dynamical. Moreover those horizons do not exist for all FLRW models. So the apparent horizon, which always exists and also local, seems to be a good choice to study thermodynamics of cosmological horizon\cite{Faraoni1, helou}.

The thermodynamics and Hawking radiation from FLRW apparent horizon has seen a lot of interest over the last few years(\cite{Cai} and references therein). The apparent horizon of FLRW spacetime is a trapping horizon of past, inner type when its Ricci scalar is positive \cite{Hayward2, Faraoni2}. Many authors have derived the temperature of this horizon using the Hamilton - Jacobi variant of Parikh - Wilczek tunnelling formalism \cite{Parikh, Vanzo, Criscienzo1, Criscienzo2, Hayward3}. Since there is no timelike Killing vector in a dynamical spacetime, particle states can not be defined in a straightforward manner. But in a special case when the dynamical spacetime has spherical symmetry, there exists a vector, named after H Kodama \cite{Kodama}, which can be used to construct a preferred time direction (hence called the {\it dynamical time}\cite{Visser}). Kodama vector is parallel to the Killing vector in the static case and there is a conserved current associated with the it. The Noether charge associated with the Kodama vector is the Misner - Sharp mass \cite{Misner, Hayward4} which is almost universally accepted as the internal energy $U$ in horizon thermodynamics. Using these properties of Kodama vector a surface gravity for the dynamical horizon can be defined \cite{Hayward2}. Also, one can construct particle states with respect to the Kodama vector.

In this paper we have applied this Hayward - Kodama prescription to the apparent horizon of FLRW spacetime in a double-null coordinate setting. Double-null coordinates though familiar for a black hole, are not so frequently used in cosmology. In these coordinates we have found the apparent horizon (trapping horizon) of FLRW spacetime using Hayward's formalism. Since the Kodama vector plays the role of time, we have constructed the positive frequency modes of the Kodama vector. The modes display discontinuity at the horizon, but this problem is tackled by considering the modes as distributions. As distributions, the modes are quite well behaved and infinitely differentiable \cite{Gel'fand} at the horizon. Then the conditional probability that the central observer receives a signal which is crossing the trapping horizon from the outside has the expected form of a Boltzmann distribution when the temperature of the horizon is $ - \kappa$ , the Kodama - Hayward surface gravity . This gives a positive temperature for the horizon, since for an inner horizon the Kodama - Hayward surface gravity is negative. We find that the horizon temperature estimated by this method matches the result obtained by Faraoni \cite{Faraoni2} and Criscienzo {\it etal} \cite{Criscienzo2} where the Hamilton-jacobi tunneling method had been employed.

This method of calculating temperature associated with a trapping horizon was previously been introduced in \cite{Chatterjee1,Chatterjee2} in the case of a black hole. It was motivated by the work of Damour and Ruffini \cite{Damour} and is found to work for different types of horizons just as well as the tunneling method.  Now it is shown here that the technique works well in a cosmological setting where it is even more required for the essentially dynamical scenario.

The plan of the paper is as follows. In section II we have discussed the general background of cosmology in double-null coordinates, the apparent horizon and the Kodama vector. In the next section we show the calculations of the modes of Kodama vector and the probability current and discuss our results in the last section. Throughout the paper the speed of light $c$, reduced Planck constant $\hbar$ and Boltzmann constant $ k_B$ are set equal to unity.

\section{Cosmology in double-null coordinates}

\subsection{Double-null coordinates for Friedmann-Lemaitre-Robertson-Walker metric}
The Friedmann-Lemaitre-Robertson-Walker line element with signature $(-,+,+,+)$ is given by,
\beq ds^2 = -dt^2 + \frac{a^2}{1-kr^2}dr^2 + R^2d\Omega^2, \eeq
where $a$ is the scale factor, $k$ is the curvature index and $R = ar$ is the invariant radius. The metric can be written in conformal time as
\beq ds^2 = - a^2(d\eta^2 - dr_*^2) + R^2d\Omega^2, \eeq where $d\eta = \frac{dt}{a}$ and $dr_* = \frac{dr}{\sqrt{1-kr^2}}$. Let us define $dx^{\pm}= d\eta \pm dr_*$ where $x^\pm$ are two null coordinates. The FLRW metric can be cast into this double-null setting as, \beq ds^2 = 2h_{+-}dx^+dx^- + R^2d\Omega^2 , \eeq
where $h_{+-}= h_{-+} = -a^2/2$; $h_{\mu\nu}$ being the metric on the 2-dimensional space spanned by $r$ and $t$, to which the spherical surfaces are orthogonal. For this metric, the Einstein's equations are given by \cite{Hayward2},
\beq
\del_{\pm}\del_{\pm} R - \del_{\pm}\log (-h_{+-})\del_{\pm} R &=& - 4\pi R T_{\pm\pm} ,\\
R \del_+\del_- R + \del_+ R \del_- R - \frac{1}{2} h_{+-} &=& 4\pi R^2 T_{+-} ,\\
R^2 \del_+ \del_- \log(- h_{+-}) - 2 \del_+ R \del_- R + h_{+-} &=& 8\pi R^2 (h_{+-} T^{\theta}_{\theta} - T_{+-}),
\eeq
where $T_{ab}$ are the components of the energy-momentum tensor.

Going back to the $(t,r)$ coordinates and considering a perfect fluid source for energy-momentum tensor, the familiar form of Friedmann equations are recovered from these equations.

We need to calculate the expansions with respect to the two null normals in this coordinates. We can construct the null normals as $l_{\pm}^{\alpha} = -g^{\alpha\mu}\del_{\mu}x^{\mp}$ and the expansions can be computed in a straightforward way, \beq \theta _+ =  \frac{1}{\sqrt -g}\partial_\alpha( \sqrt-g l_+^\alpha) = \frac{2}{a^2}\frac{2}{R}\partial_+ R ,\eeq and
\beq \theta _- = \frac{1}{\sqrt -g}\partial_\alpha( \sqrt-g l_-^\alpha) = \frac{2}{a^2}\frac{2}{R}\partial_- R .\label{exp}\eeq

\subsection{Apparent horizon for FLRW spacetime}
In standard cosmology two types of horizons are well known, the particle horizon and the event horizon. At a given time $t$,  the particle horizon for an observer located at $r=0,$ is the boundary between the worldlines of events that can be seen and the events that can not be seen. It is a past horizon. The event horizon is the proper distance to the most distant event that the observer will ever see. So it is a future horizon. Both of these horizons, which are null surfaces, are defined globally i.e one has to know the entire past or future evolution of the universe in order to define these horizons. So for our purpose these concepts are not very useful. Instead we consider the apparent horizon of FLRW spacetime which can be defined locally and unlike event horizons, exists in all FLRW models.

\subsubsection{\bf Definition}
The apparent horizon of FLRW spacetime is a trapping horizon of past, inner type. Following Hayward \cite{Hayward2}, a past  inner trapping horizon (PITH) is defined as follows.
 A 3-dimensional hypersurface $\Delta$ in 4-dimensional FLRW spacetime is said to be a PITH if
 \begin{itemize}
 \item It is foliated by marginal 2-spheres on which expansion of one null normal is zero (here $\theta_-$) and the expansion of other null normal is positive definite(here $\theta_+$).
     \item The directional derivative of $\theta_-$ along the null normal $l^{+}$ (i.e, $\lie_{l^+}\theta_- $) is positive definite, \end{itemize}
         where the double-null foliation is adapted to the marginal surfaces. For FLRW spacetime in double-null coordinates using equations (\ref{exp}), the PITH is located at $\theta_- = 0$. Also the second requirement becomes, $\del_+ \theta_- > 0$ on $\Delta$.

\subsubsection{\bf Condition for timelike PITH} To see the nature of the trapping horizon, let us consider a tangential vector field on $\Delta$ $ t^\alpha = l_-^\alpha + h l_+^\alpha$ for any smooth function $h$. The norm of $t$ is $t^\mu t_\mu = -h\frac{4}{a^2} .$ This shows the nature of PITH depends on the sign of $h$ i.e, $\Delta$ is timelike if $h > 0$, spacelike if $h < 0$ and degenerate if $h = 0$. Since $\theta_- = 0$ on $\Delta$, we have $\lie_{t} \theta_-\mid_\Delta = 0.$ This implies $\del_-\theta_- = -h \del_+ \theta_- .$ Considering a perfect fluid with equation of state $p = w \rho ,$ and using Friedmann's equations we get, \beq \del_- \theta_- = 4\pi G (\rho + p) \\
\del_+ \theta_- = \frac{4 \pi G}{3} (1 - 3w)\rho .\eeq and \beq h = \frac{3(1 + w)}{(1 - 3w)} . \eeq
\begin{itemize}
\item The trapping horizon $\Delta$ is timelike when $ -1 < w < \frac{1}{3} .$
\item $\Delta$ is spacelike when $ w < -1$ or $ w > \frac{1}{3} .$
\item $\Delta$ is degenerate if $w = -1 . $
\end{itemize}

\subsection{The Kodama vector}
For a spherically symmetric dynamical spacetime with line element
\beq ds^2 = h_{ij}dx^i dx^j + r^2d\Omega^2 , \eeq
where $r=r(x^{i})$ is the proper radius, the Kodama vector is,
\beq K^i = \epsilon^{ij}\del_{j}r ,\eeq
with $ K^\theta = K^\phi = 0,$ where $\epsilon^{ij}$ denotes the volume form associated with the two-metric $h_{ij}$ \cite{Kodama}. So, the Kodama vector lies in the plane orthogonal to the sphere of symmetry. The importance of Kodama vector is that it generates a preferred flow of time in dynamical spacetime (spherically symmetric). Using Einstein's equations it can be shown that the vector field $ J^a = G^{ab}K_b$ is divergence free where $G^{ab}$ is the Einstein tensor \cite{Kodama,Visser}. This implies that even though the spacetime is completely dynamical, $J^a$ is the locally conserved energy flux. So the Kodama vector provides a preferred timelike direction. Also, it is shown in \cite{Hayward4}, that the Noether charge associated with the Kodama current $J^a$ leads to the Misner-Sharp energy \cite{Misner} of the spacetime.
For FLRW spacetime in dual-null coordinates, the Kodama vector is given by,
\beq K = \frac{2}{a^2}\{\del_{+}R\del_- - \del_{-}R\del_+\}. \eeq

\subsection{The surface gravity} The Kodama-Hayward surface gravity $(\kappa)$ for spherically symmetric dynamical spacetime is defined using the Kodama vector $(K)$ as\cite{Hayward2},
\beq K^i\grad_{[j}K_{i]} = \kappa K_j .\eeq
Another equivalent definition of $\kappa$ is
\beq \kappa = \frac{1}{2}\Box_{(h)}R = \frac{1}{2\sqrt {-h}}\del_{\mu}(\sqrt{-h}h^{\mu\nu}\del_{\nu}R). \eeq
 For FLRW spacetime in dual-null coordinates, it takes the form,
 \beq \kappa = -\frac{2}{a^2}\del_+\del_- R. \eeq
The negative sign in $\kappa$ appears due to the inner nature of the trapping horizon \cite{Criscienzo2}.

\section{Positive frequency modes of Kodama vector and probability current}
To calculate the temperature associated with the trapping horizon, we will utilize the Kodama vector. Since it gives a preferred timelike direction, we can construct well behaved field modes both inside and outside the horizon. Then temperature is found by equating the conditional probability that a signal reaches the observer when it is coming from outside the horizon to the Boltzmann factor.

One can find positive frequency eigenmodes of the Kodama vector using,
\beq i K Z_\omega = \omega Z_\omega.\label{mode} \eeq $Z_\omega$ are the eigenfunctions corresponding to $\omega (>0)$. To simplify the calculations let us change coordinates $ (x^+, x^-)$ to $ (y = x^+, R)$, so that,
\beq \del_+ = \del_y + \del_+ R\del_R \\
\del_- = \del_-R\del_R . \eeq With these transformations, we get,
\beq K Z_\omega = -\frac{2}{a^2}F(R)\del_y\tilde{Z_\omega},\eeq where $ Z_\omega(x^+,x^-) = \tilde{Z}_\omega(y,R) $ and $F(R) = \del_- R$. So equation (\ref{mode}) becomes,
\beq \del_y \tilde{Z_\omega} = i \frac{a^2}{2}\frac{\omega}{F(R)}\tilde{Z_\omega}. \eeq  Integrating and transforming back to old coordinates the above equation gives, \beq Z_\omega = C(R)\exp\Big(i\omega\int_R \frac{dx^+}{\frac{2}{a^2}\del_- R}\Big),\label{int}\eeq where $C(R)$ is an arbitrary smooth function of $R$ (any smooth function of $R$ is a zero mode of Kodama vector $K$) and the subscript $R$ under the integral sign denotes R is kept fixed while doing the integration. To evaluate the integral in (\ref{int}), we multiply the numerator and denominator by $(\del_+ \theta_-)$ and use the fact that for any fixed R surfaces, $\del_+ \theta_- = -2\kappa/R$. So the integral in(\ref{int}) becomes
\beq i\omega\int_R \frac{dx^+ \del_+ \theta_-}{\frac{2}{a^2}\del_- R \del_+ \theta_-} = -i\omega\int_R \frac{d\theta_-}{\frac{2\kappa}{R}\frac{2}{a^2}\del_- R }= -i\omega\int_R \frac{d\theta_-}{\kappa\theta_- }. \eeq  We now make the only assumption required in this calculations namely during the evolution of the dynamical horizon $\kappa$ is a slowly varying function in some small neighbourhood of the horizon (the zeroth law takes care of it on the horizon but we assume it to hold in a small neighbourhood of the horizon) so that we can take $\kappa$ out of the integral sign and the result is,

\beq Z_\omega=C(r)\begin{cases} \theta_-^{-\frac{i\omega}{\kappa}} &
\text{for}\;\theta_->0\\
(|\theta_-|)^{-\frac{i\omega}{\kappa}} & {\rm for}\;
\theta_-<0.\end{cases}\label{tmode}\eeq

where the spheres are not trapped "outside the trapping horizon" $(\theta_- > 0)$ and fully trapped "inside" $(\theta_- < 0)$. It seems that the modes are defined both inside and outside the horizon but not on the horizon. Now we have to keep in mind that the field modes (\ref{tmode}) are not ordinary functions, but are distribution-valued, and as distributions (or generalized functions) these are perfectly well behaved on the horizon. Following \cite{Gel'fand} the appropriate distribution for the positive frequency modes are,
\beq Z_\omega = \lim_{\epsilon \to 0} C(R)(\theta_{-} + i\epsilon)^{-\frac{i\omega}{\kappa}}
= \left\{ \begin{array}{ll}
           C(R)\left(\theta_{-}\right)^{-\frac{i\omega}{\kappa}} & \mbox{ for $\theta_{-} > 0$,} \\  C(R)|\theta_{-}|^{-\frac{i\omega}{\kappa}}
e^{\frac{\pi\omega}{\kappa}} & \mbox{ for $\theta_{-} < 0$. }
            \end{array}
\right.
\eeq
The complex conjugate modes are given by,
\beq Z_{\omega}^* = \lim_{\epsilon \to 0} {C(R)}^*(\theta_{-} - i\epsilon)^{\frac{i\omega}{\kappa}}
= \left\{ \begin{array}{ll}
           {C(R)}^*\left(\theta_{-}\right)^{\frac{i\omega}{\kappa}} & \mbox{ for $\theta_{-} > 0$,} \\  {C(R)}^*|\theta_{-}|^{\frac{i\omega}{\kappa}}
e^{\frac{\pi\omega}{\kappa}} & \mbox{ for $\theta_{-} < 0$. }
            \end{array}
\right.
\eeq
Let us now calculate the probability current associated with these modes along the direction of the Kodama vector $K$ across the horizon
\beq \varrho(\omega) = \frac{i}{2}[Z_\omega^* K Z_\omega - Z_\omega K Z_\omega^*] = \omega Z_\omega^* Z_\omega .\eeq
A straightforward calculation gives, apart from a function of $R$,
\begin{align}
\varrho(\omega)&=\omega(\theta_- - i\epsilon)^{\frac{i\omega}{\kappa}}
(\theta_- + i\epsilon)^ {-\frac{i\omega}{\kappa}}.\nn
&=\begin{cases} \omega & \text{for}\;\theta_->0\\
\omega e^{\frac{2\pi\omega}{\kappa}} & \text{for}\;\theta_-<0.\end{cases}
\end{align}

The conditional probability that a signal reaches the central observer when it is crossing the horizon from the outside i.e the incoming probability is given by,
\beq
P_{(emission|incident)} =  \frac{P_{(emission{\cap}incident)}}{P_{(incident)}} = \frac{\omega e^{\frac{2\pi\omega}{\kappa}}}{\omega} =  e^{\frac{2\pi\omega}{k}}.
\eeq
Comparing this with the Boltzmann factor, $e^{-\beta \omega}$ the temperature of the apparent horizon is found to be $-\kappa/2\pi$ which is positive since $\kappa$ is negative.

\section{Discussions}
In this paper we have discussed a different approach of calculating horizon temperature for FLRW cosmological models. Our derivation depends on two crucial points. The first one is the geometric structure of the background spacetime. Since FLRW spacetime is spherically symmetric, the Kodama vector and the Misner-Sharp energy can be defined uniquely and this is essential for our calculations. For a non-spherical geometry, a Kodama-like vector field is not known yet and its not obvious how our formalism can be extended to these type of geometries. The second important point is the assumption of slow variation of surface gravity across the trapping horizon. Our calculation shows that the temperature of the trapping horizon is $-\kappa/2\pi$ when $\kappa$ is a slowly varying function in the vicinity of the horizon. This is a rather simplifying assumption which enables us to find the mode solutions in a closed form. If it was not so, the general method would stand, but it would have been rather difficult to identify $-\kappa/2\pi$ as temperature. At this point, we should also mention that in this formalism the Hayward - Kodama surface gravity emerges as the temperature associated with the horizon naturally, although several other definitions of surface gravity associated with a dynamical horizon are available in the literature. In this regard our result agrees with \cite{Criscienzo2, Faraoni2}, where horizon temperature was calculated using Hamilton - Jacobi tunneling method.

 As was mentioned earlier, this formalism was introduced in \cite{Chatterjee1,Chatterjee2} and applied to the stationary and dynamical black hole horizons including extremal ones. The present work is the first attempt to find horizon temperature in a cosmological scenario using this formalism. The main advantage of this formalism is that the calculations are exact and no approximations are used in obtaining the field modes whereas the widely used tunneling method depends heavily on the WKB approximation \cite{Parikh,Vanzo}. It remains to be seen how this method works for other diffeomorphism invariant theories of gravity.

\section{Acknowledgements}

We would like to thank Tanima Duary for useful discussions. B.C. also acknowledges DST Project Grant No. SR/WOS-A/PM-78/2017 under the DST Women Scientists Scheme A.


\begin{thebibliography}{100}
%
\bibitem{Bardeen}
  J.~M.~Bardeen, B.~Carter and S.~W.~Hawking,
  Commun.\ Math.\ Phys.\  {\bf 31}, 161 (1973).
%
\bibitem{Bekenstein}
  J.~D.~Bekenstein,
  Phys.\ Rev.\  D {\bf 7}, 2333 (1973).
%
  \bibitem{Hawking}
  S.~W.~Hawking,
  Commun.\ Math.\ Phys.\  {\bf 43}, 199 (1975)

%
\bibitem{Gibbons}
    W. Gibbons and S. W. Hawking,
    Phys.\ Rev.\  D {\bf 15}, 2738 (1977).
%

\bibitem{Ashtekar1}
A. Ashtekar and B. Krishnan,
Living Rev. Relativity {\bf 7}, 10 (2004).
%

\bibitem{Faraoni1}
Valerio Faraoni,
    Lecture\ notes\ in\ physics\ (volume 907) -  (2015, Springer).


\bibitem{Hayward1}
  S.~A.~Hayward,
  Phys.\ Rev.\  D {\bf 49}, 6467 (1994).
%
\bibitem{Hayward2}
  S.~A.~Hayward,
  Class.\ Quant.\ Grav.\  {\bf 15}, 3147 (1998)
  [arXiv:gr-qc/9710089].
%
\bibitem{Booth}
  I.~Booth and S.~Fairhurst,
  Phys.\ Rev.\ Lett.\  {\bf 92}, 011102 (2004)
  [arXiv:gr-qc/0307087].
%

\bibitem{Krishnan}
 A.~Ashtekar and B.~Krishnan,
  Phys.\ Rev.\ Lett.\  {\bf 89}, 261101 (2002)
  [arXiv:gr-qc/0207080]; Phys.\ Rev.\  D {\bf 68}, 104030 (2003)
  [arXiv:gr-qc/0308033].
%
\bibitem{Ashtekar2}
  A.~Ashtekar, C.~Beetle, O.~Dreyer, S.~Fairhurst, B.~Krishnan, J.~Lewandowski and J.~Wisniewski,
  Phys.\ Rev.\ Lett.\  {\bf 85}, 3564 (2000)
  [arXiv:gr-qc/0006006].

\bibitem{helou}
P. Binetruy and A Heou, Class. Quantum Grav., {\bf 32}, 205006(2015).


\bibitem{Cai}
 Rong-Gen Cai and Sang Pyo Kim,
 J.\ High\ Energy \ Phys. {\bf 02}, 050 (2005). A. B. Nielsen and D.-H. Yeom, Int.\ J. \ Mod.\ Phys.\ A {\bf 24},
5261 (2009); M. Akbar and R.-G. Cai, Phys.\ Lett.\ B {\bf 635}, 7
(2006); R.-G. Cai and L.-M. Cao, Phys.\ Rev.\ D {\bf 75}, 064008
(2007); M. Akbar and R.-G. Cai, Phys.\ Lett.\ B {\bf 648}, 243
(2007); R.-G. Cai and L.-M. Cao, Nucl.\ Phys.\ B {\bf 785}, 135
(2007); A. Sheykhi, B. Wang, and R.-G. Cai, ibid.  B {\bf 779}, 1
(2007); Phys.\ Rev. \ D {\bf 76}, 023515 (2007); R.-G. Cai, L.-M.
Cao, and Y.-P. Hu, J.\ High \ Energy \ Phys. {\bf 08} 090 (2008);
R.-G. Cai, Prog.\ Theor.\ Phys.\ Suppl.{\bf 172}, 100 (2008);R.-G. Cai,L.-M. Cao and Y. -P. Hu, Class.\ Quantum.\ Grav {\bf 26}, 155018
(2009); Tao Zhu et al. Int.\ J. \ Mod.\ Phys.\ D {\bf 19} 159, (2010).
%

 \bibitem{Faraoni2}
  Valerio Faraoni,
  Phys.\ Rev.\  D {\bf 84}, 024003 (2011).
%

\bibitem{Parikh}
  M.~K.~Parikh and F.~Wilczek,
  Phys.\ Rev.\ Lett. {\bf 85},  5042-5 (2000).
  %
\bibitem{Vanzo}
  L.~Vanzo, G.~Acquaviva and R.~Di.~Criscienzo
  Class.\ Quantum \ Grav. {\bf 28}  183001 (2011).
%
\bibitem{Criscienzo1}
  R.~Di.~Criscienzo, M.~Nadalini, L.~Vanzo, S.~Zerbini and G.~Zoccatelli,
  Phys.\ Lett.\ B {\bf 657}  107-11 (2007).
%
\bibitem{Criscienzo2}
  R.~Di Criscienzo, S.~A.~Hayward, M.~Nadalini, L.~Vanzo and S.~Zerbini,
  Class.\ Quant.\ Grav.\  {\bf 27}, 015006 (2010)

%
\bibitem{Hayward3}
  S.~A.~Hayward, R.~Di Criscienzo, L.~Vanzo, M.~Nadalini and S.~Zerbini,
  Class.\ Quant.\ Grav.\  {\bf 26}, 062001 (2009)
  [arXiv:0806.0014 [gr-qc]].
%
\bibitem{Kodama}
  H.~Kodama,
  Prog.\ Theor.\ Phys.\  {\bf 63}, 1217 (1980).
%
\bibitem{Visser}
G. Abreu and M. Visser,
 Phys.\ Rev.\ D {\bf 82}, 044027(2010).
 %

  \bibitem{Misner}
C.W. Misner and D. H. Sharp,
Phys.\ Rev.\ {\bf 136}, B571,(1964).
%
\bibitem{Hayward4}
  S.~A.~Hayward,
  Phys.\ Rev.\  D {\bf 53}, 1938 (1996).
  %
\bibitem{Gel'fand} I.M Gel'fand and G.E Shilov  {\it Generalized Functions V1} {(Academic Press)} (1964)

%
\bibitem{Chatterjee1}
  B.~Chatterjee and A.~Ghosh,
  J.\ High\ Energy\ Phys.\ {\bf JHEP 1204},\ 125 (2012).

  %
  \bibitem{Chatterjee2}
  A Chatterjee, B Chatterjee, A Ghosh,
 Phys.\ Rev.\  D {\bf 87},\ 084051 (2013),
 [arXiv:1204.1530 [gr-qc]]

 %
  \bibitem{Damour}
  T.~Damour and R.~Ruffini,
  Phys.\ Rev.\  D {\bf 14}, 332 (1976).
%

\end{thebibliography}
\end{document}